\documentclass{article}
\usepackage{arxiv}
\usepackage[utf8]{inputenc} 
\usepackage[T1]{fontenc}    
\usepackage{geometry}
\usepackage{pifont}
\usepackage{tabu}
\usepackage{float}
\usepackage{microtype}      
\usepackage{graphicx}
\usepackage[backend=bibtex,
  citestyle=numeric-comp,
  bibstyle=numeric,
  sorting=none,
  indexing=true]{biblatex}

\newcommand{\cmark}{\ding{51}}
\newcommand{\xmark}{\ding{55}}
\title{A Survey of Task-Based Machine Learning Content Extraction Services for VIDINT}

\author{
  Joshua Brunk \\
  University of Maryland College Park\\
  Dept. of Electrical and Computer Engineering\\
  \texttt{jbrunk@umd.edu} \\
  \And
  Nathan Jermann \\
  Franciscan University \\
  Dept. of Mathematics and Computer Science\\
  \texttt{njermann001@student.franciscan.edu} \\
  \And
  Ryan Sharp \\
  University of Maryland Baltimore County \\
  Dept. of Computer Science and Electrical Engineering\\
  \texttt{jsharp2@umbc.edu} \\
  \And
  C.D. Hoover, Ph.D. \\
  Elder Research, Inc.\\
  \texttt{hoover@datamininglab.com} \\
}
\date{}
\begin{document}

\maketitle

\begin{abstract}
This paper provides a comparison of current video content extraction tools with a focus on comparing commercial task-based machine learning services. Video intelligence (VIDINT) data has become a critical intelligence source in the past decade. The need for AI-based analytics and automation tools to extract and structure content from video has quickly become a priority for organizations needing to search, analyze and exploit video at scale.  With rapid growth in machine learning technology, the maturity of machine transcription, machine translation, topic tagging, and object recognition tasks are improving at an exponential rate, breaking performance records in speed and accuracy as new applications evolve. Each section of this paper reviews and compares products, software resources and video analytics capabilities based on tasks relevant to extracting information from video with machine learning techniques.
\end{abstract}


\section{Introduction}

In recent years{\footnote{This technology assessment was completed in 2019 by its authors working as data science interns at Elder Research. Due to the rapidly changing landscape of technologies and providers referenced herein, pricing and availability of these VIDINT tools are expected to change by the time of its publication.}, we've seen exponential growth in data, with one estimate that 90 percent of worldwide data was generated in the last two years alone \cite{forbes}. This surge in data is also present in the most inclusive media format, video data, with both television and internet sources contributing. For example, it is estimated that there are 400+ channels broadcast internationally that provide 24-hour news service \cite{wikiTV} with thousands more that provide 8 or less hours a day \cite{journalism}. For reference an hour of 720p digital video is around 1.5 GBs of data. YouTube currently ingests around 880 TB of data per day \cite{brandwatch} from 91 countries in 80 languages \cite{ytpress}. With all this data available, it becomes a challenge to obtain meaningful information from videos and broadcasts without actually watching them. The only information searching is able to reveal are video titles. 

The purpose of this document is to summarize information about the current state of the art in video analytics tools.  Specifically, we will focus on comparing tools and techniques in the following sub-problems: machine transcription (speech-to-text), machine translation, segmentation, topic tagging, optical character recognition (OCR), and general object recognition. These tasks enable smarter extraction of information from video, which can improve searching videos for information for a myriad of exploitation applications. Our analysis for each solution falls into three main topics: capabilities, performance (as best as possible), and cost. Large vendors are offering mature cloud based services for these tasks, with more features and accuracy improvements being released regularly.

When exploring the state of the art, we were primarily interested in the application of enriching worldwide broadcast news videos. In 2017, broadcast TV was found to be a more consumed media than radio, streaming video, and print news periodicals \cite{emarketer}. Over the course of this paper we will provide examples of enrichment, with most of our enhancements focused on the characteristics of newscasts, such as talking, on-screen text, and imagery relevant to the subject matter. For illustrative purposes, an example newscast screenshot and transcript is given below (Fig. \ref{news1}):

\begin{figure}[H] 
  \centering
  \includegraphics[width=13cm]{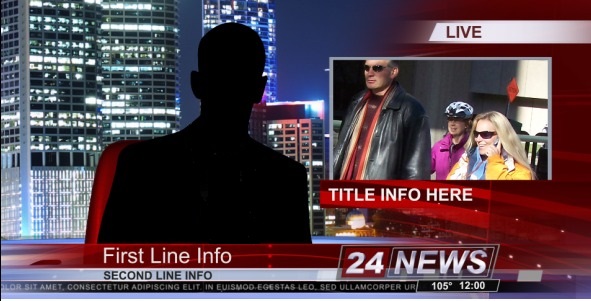}
  \caption{News Video Footage Example \cite{videohive}}
  \label{news1}
\end{figure}

\paragraph{Transcript from news report: }
\textbf{Source: }
\textit{"Acabamos de recibir informes de que el Sr. K y su esposa, la Sra. K, han cerrado exitosamente un acuerdo con el fabricante internacional de piezas, ACME Corporation.}  \\
\textbf{English Translation: } 
\textit{"We've just received reports that Mr. K and his wife Mrs. K have successfully closed a deal with international parts manufacturer, ACME Corporation."}

\section{Machine Transcription}

The process of Machine Transcription is extracting human speech from an audio source and converting it to text in the source language. Most videos include some form of verbal communication, which is usually the primary source of information in the video. The goal of transcription is to convert the audio into text from which we can extrapolate key information. While applicable for newscasts, transcription may not be useful in many videos that do not have talking, such as surveillance video. When the content type is unknown, an additional layer that focuses on speech detection would be useful. For this paper we will focus on spoken communication being present in a video, and it being the primary source of information from videos.

Current out-of-the-box transcription services from companies such as Google, Amazon, and IBM have remarkably high accuracy for a variety of languages, dialects, and accents; however, there is variation in performance when the type of audio data changes. For example, some transcription services prefer a certain sampling rate, usually 16 kHz, and have built separate models for 8 kHz audio. The word error rate (WER) on ideal audio for some of the out-of-the-box solutions can be as low as 5\% on average \cite{wer}, however, there are issues that affect performance markedly. For example, performance takes a significant hit when there is cross talk (people talking concurrently), or when there is a consistent ambient noise from the environment from a source such as being inside a moving car.   

Most transcription services provide metadata alongside their transcriptions. This generally includes speaker recognition and labeling, alternative transcriptions, confidence scores, and time stamps for words. Other standard advantages of transcription services include channel identification (left or right stereo channels), real time applications, custom library integration, and inappropriate content filters. 

The sample newscast transcription could include timestamps for a series of text: \textbf{\textit{Sra. K, han cerrado exitosamente un acuerdo}} (time: 01:13 - 01:15), which after finding topic segments (see Section  \ref{textsec}) would allow lookup on detected topics to play a key segment of the video. The transcription output may also provide alternate words, given high confidence in the other word: \textbf{\textit{acuerdo - 96\%, recuerdo - 83\%}}. Some transcription may also include key tokens for a segment of the audio: \textbf{\textit{Sr. K, Sra. K, Dataland, ACME, Umbrella Co.}} (01:11 - 01:21).

\subsection{Amazon Transcribe}

Amazon's API allows users to submit video and audio data in a variety of formats and automatically chooses the best Amazon transcription model, whereas other services will have a user select the model they want to use and format the audio appropriately. Transcribe also supports both real-time and asynchronous transcription of audio. Audio sampling is supported from 8 – 48 kHz, with MP3, FLAC, or PCM encoding. Amazon Transcribe supports 8 languages, including dialects for several of them, and allows user supplied custom word libraries - although users cannot supply labeled training data, as the model is managed and trained by Amazon. In regards to integration, transcribe works with .NET, Go, Java, JavaScript, Python, C++, and Ruby SDKs. One technical drawback to Amazon Transcribe is a smaller batch size of 4 hours, or 2 GB.  It should also be noted that Amazon Transcribe also restricts batch size for audio data not stored within an Amazon S3 bucket. 

Amazon Transcribe's free tier provides 60 minutes of standard audio transcription every month, for the first 12 months. Their regular pricing is \$0.024 per minute, with price reductions for bulk orders\footnote{All pricing in this assessment is reported at the time of writing circa 2018-2019. Prices are expected to change over time.}  Both the free transcription amount and the standard pricing are below average among paid translation solutions \cite{amztsc}.

\subsection{Google Cloud Speech-to-Text}

Google Cloud Speech-to-Text has three main request methods for speech recognition: Synchronous Recognition, which processes small files (up to one minute) all at once; Asynchronous Recognition, which processes large files (480 minutes), and allows the user to pull interim results; and Streaming Recognition, which can perform real time transcription with much faster interim result polling. Cloud Speech-to-Text's API supports audio with 8 - 48 kHz sampling rates with MP3, FLAC, PCM, u-LAW, AMR, or OPUS encoding; FLAC and Linear PCM provide the best transcriptions given they're lossless. Drawing from their translation software, Google Cloud Speech-to-Text is capable of transcription and translation across 120 languages and variants. Alongside the largest available batch size (up to 8 hours of audio), Google Cloud Speech-to-Text provides specialized models for certain transcription tasks. An example of this would be their "Phone call" version of their transcription service, which has undergone extensive training specifically on phone audio (8 kHz). Google Cloud Speech-to-Text also offers the ability to supply personalized dictionaries to strengthen transcription. Although, much like Amazon, it is designed to work closely with Google Cloud in order to send and receive transcriptions. This can make it especially difficult when seeking to incorporate this software with other services.

Cloud Speech-to-Text also allows 60 minutes of free transcription service, but only for first time use, after that base pay ranges from \$.024 - \$0.016 with additional costs for more specialized models \cite{ggtsc,gaibtsc}.

\subsection{IBM Watson Speech-to-Text}

Watson provides three separate interfaces to cope with different transcription demands. Their WebSocket interface specializes in establishing a continuous high-speed connection with their service, allowing passage of up to 100 MB of audio per batch, in real time. The Synchronous HTTP interface is for basic HTTP calls, this also limits the maximum batch size to 100 MB. Finally, the asynchronous HTTP interface allows the user to transcribe non-blocking calls to their service, with a maximum batch size of 100 MB. Watson supports 8 and 16 kHz sampled audio in MP3, FLAC, VP8, VP9, PCM, or u-LAW encoding, with compression capabilities that allow more processing ability for the same price. Watson is one of the most customizable transcription services available on the market. It grants use of not only custom word and grammar libraries, but entire custom models as well, and there are options for spotting key words throughout a transcription. Where Watson falls short is with language, as not all features or models are available for all languages, and quality drops for specific languages. There are also reported accuracy problems when multiple speakers are present in an audio file.

IBM has the least expensive pricing model. Their free model provides 100 minutes per month, for an unlimited period of time. And their base model costs \$0.01 - 0.02 per minute, with no additional fees for extra features. Watson doesn’t need additional software to operate, such as IBM cloud storage \cite{ibmtsc}.

\subsection{DeepGram}

DeepGram offers Speech-to-Text in their base model, and allows clients to customize a model with DeepGram directly. DeepGram includes models optimized for phone calls, meetings with significant cross-talking, and a general model. Language support is very limited as only English, Spanish, and Korean are supported. DeepGram supports most audio formats such as MP3, FLAC, AAC, as well as varying sampling rates, as long as bit-rate is decent; Deepgram also allows direct uploads of videos and will strip the audio themselves. Like other vendors, DeepGram offers real-time and asynchronous transcription calls, and can also deploy models on site. DeepGram features the ability to search for sections of audio directly, readjust timestamps, and automatically detects phoneme patterns.

Pricing is done on a customer to customer basis, with models tailored for specific tasks; exact prices are determined after a consultation \cite{deepgram}.

\subsection{Vendor Comparison}

\begin{table}[H]
\tabulinesep=1mm
\begin{tabu} to \linewidth{ X[1.2,l,m] X[1,C,m] X[1,C,m] X[1,C,m] X[1,C,m] }
\tabucline[1.5pt]{-}
{} & \textbf{Amazon Transcribe} & \textbf{Google Cloud Speech-to-Text} & \textbf{IBM Watson Speech-to-Text} & \textbf{DeepGram} \\
\hline \hline
\textbf{Base Price} \newline (per minute) & \$0.024 & \$0.016-\$0.024 & \$0.02 & Per customer basis \\
\textbf{Free Tier} \newline (minutes per month) & 60 (first 12 months) & 60 (one month only) & 100 & N/A \\
\textbf{Audio Sampling Rate} \newline (khz) & 8 - 48 & 8 - 48 & 8 or 16 & 8 - 48 \\
\textbf{Custom Vocabulary} & \cmark & \cmark & \cmark & \cmark \\
\textbf{Batch Size} & 2 GB & ($\sim$4 GB) & 100 MB & N/A \\
\textbf{Accepted Formats} & MP3, FLAC, PCM & MP3, FLAC, PCM, u-LAW, AMR, OPUS & MP3, FLAC, VP8, VP9, PCM, u-LAW & MP3, FLAC, AAC \\
\tabucline[1.5pt]{-}
\tabuphantomline
\end{tabu}
\caption{Transcription Vendor Comparison}
\end{table}

\subsection{Supplementary Options}
\subsubsection{Dragon Speech Recognition Software (Nuance)}

Utilizing Nuance's own deep learning model, Dragon specializes in personal voice transcription. Known for being extremely accurate, Dragon works with accents, custom voices, and provides many different models for separate use cases. However, Dragon should not be used for transcribing random audio files as it was created with the intent of being a personal transcription service since Nuance’s learning model trains specifically with a user’s voice. Minimum file sampling rate is 16 kHz and the software can work with many files types, but traditionally takes in audio directly from a microphone. Dragon software is a one-time purchase that costs between \$150 - \$300 \cite{dragon}.

\subsubsection{Kaldi Speech Recognition Toolkit (Open Source)}

Kaldi is the best-known open source software framework used for speech recognition. Written in C/C++ but available in Python and Bash formats, Kaldi provides a base model which has to be trained and tested before use, but many pre-trained models already exist since it is open source. Some of these models include the ASpIRE Chaine Model (English), CVTE Mandarin Model, SRE16 Xvector Model (English) and others, some of which have WER's as low as 3\% \cite{kaldi}. Kaldi also supports use of deep neural networks and is continually updated for increased accuracy and ease of use. Kaldi requires user data to train the model and adapt to specific domains, but since it is free to use it can easily offset the upfront costs of High Performance Computing (HPC) machines \cite{kaldigit}.

\subsubsection{Mozilla \& DeepSpeech (Open Source)}

Mozilla DeepSpeech is Mozilla’s open source Speech-to-Text model that utilizes Google's TensorFlow\footnote{TensorFlow is an open source code library optimized for machine leaning functions, with support for the Keras API \cite{tensorflow}.} \cite{tensorflow} as well as Baidu's DeepSpeech Framework. There are a few pre-trained models on limited languages (primarily English) that also have restrictions on sampling rate, file format, and bit-rate (16 kHz, WAVE, 16-bit respectively). It is compatible with CUDA dependencies to take advantage of GPU computing. Baidu's DeepSpeech \cite{baidu} can be trained entirely from scratch in multiple languages, but, like Kaldi, will require more upfront costs and time to train the model \cite{mozilla}.

\section{Machine Translation}

Machine Translation is the process of taking written or audio communication and converting it from one language to another. Today, with over 6,000 spoken languages worldwide, many countries are becoming more involved in the digital world. As seen in Fig. \ref{graph_0}, a sample of the worlds 24-hour television news broadcasts alone constitutes a collection of 24 separate countries, the majority of whom contain individual languages. Translation would primarily be a step that is done after transcription, which retains the source language as translation is usually (but not always) done on text data. Currently, the machine translation of audio data relies on pre-processing the data (using speech-to-text or other schemes), and then performing the translation on the processed data. Performance varies significantly from language to language, due to the availability of labeled data; the more widely a language is used, the more data is generally available for it. Model performance is also effected when translating by languages in different classes. For example, romance languages are typically more similar in structure and have more direct translation mappings to one another. 

\begin{figure}[H] 
  \centering
  \includegraphics[width=\linewidth]{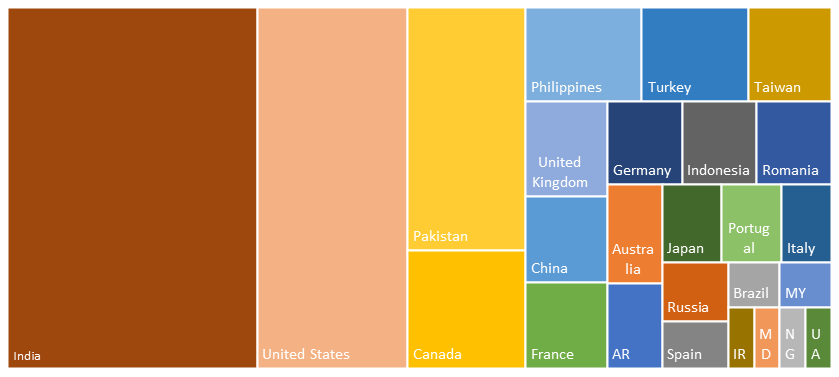}   
  \caption{24-Hour Television News Broadcasts by Country}
  \label{graph_0}
\end{figure}

Few vendors provide comprehensive performance statistics about their models, mostly for two reasons. First, the underlying models are not static and are constantly being re-trained by new data as it becomes available, altering performance. Second, the performance varies based on the quality of the input data. The worse or more specialized the input, the poorer the performance.  

As machine translation has evolved, vendors have kept legacy translation models available for use, but they generally perform worse than their new, typically deep neural network models. These models are retained for discussing improvements in performance, but may also be the only available model for a language pair. Additionally, most translation services provide features to modify or fine-tune the models, either through user supplied dictionaries of domain specific terms, or through a user supplied corpus used to modify models for a specific language pair. Like transcription, vendors do not provide the underlying implementation details of their machine learning models, and it is suspected that they are in a perpetual state of iterative improvement and retraining based on the availability of new data sources.  

Output could look like the following: 
\begin{itemize}
   \item \textbf{\textit{"We've just received reports that Mr. K and his wife Mrs. K have successfully closed a deal with international parts manufacturer, ACME Corporation. This will reduce shipping costs from their Dataland factories, and strengthen relations between Umbrella Co. and ACME."}}
   \item \textbf{\textit{"We have just received information that the MK and his wife, Ms. K, successfully celebrated a contract with the international spare parts manufacturer ACME Corporation, which will reduce the cost of shipping at their Dataland plants and strengthen the relationship between Umbrella Co. and Acme."}}
\end{itemize}
Two sample translations are included as it is common for variation between tools though usually without losing the meaning of the text. Evaluating the two translations above, there are different adjectives, such as ACME being a \textit{'spare'} parts manufacturer, as well as the second translation combined the two sentences into one larger sentence. While this example doesn't have a significant change of meaning, accuracy is very important for the task of translation so as not to skew the information.

\subsection{Amazon Translate}

Amazon Translate currently supports 25 languages with automatic language detection, but has plans to add more languages in the future. Amazon strictly uses modern LSTM neural networks in all of its translations, which provides for more consistent results in the languages they support than other services that may rely on legacy models. There are several language pairs among the 25 supported languages that are not currently supported (Korean – Hebrew, for example). Similar to other services, there is the additional feature of being able to provide translated word pairs to improve specific language pair models. 

Amazon translate is designed to be integrated into Amazon Web Service's (AWS) overall ecosystem, with all data sent between any application and Amazon’s translation service being stored with at-rest encryption within the AWS region where the service is being used. This also means that administrators can control access to Amazon Translate through AWS Identity and Access Management (IAM) permissions policy. 

Amazon’s translation pricing costs \$15 per million characters (25\% cheaper than Google) and additionally includes 2 million free characters per month for the first year (four times the amount of free characters per month as Google’s free tier) \cite{amztrans}.

\subsection{Google Cloud Translation}

As with transcription, Google boasts the most language support among major vendors, supporting 104 languages. Additionally, for any supported language, all language pairs are available. In general, Google’s translations are very accurate, but the accuracy of any specific language pair is unique. An English to French translation, for example, may perform better than an English to Zulu translation. There are not currently any overall performance metrics available across all of Google’s supported language pairs.  

Some notable features of Google’s translation offering are language detection (with confidence levels), integration within Google’s cloud services ecosystem, and the ability to supply data to models to further train specific models. For language detection, Google’s API provides a request function that returns a list of detected languages along with confidence levels for those languages. Google offers their pre-trained translation model as well as their Beta AutoML \cite{automl} service, which enables training a domain-specific model at the cost of a smaller feature set.

Google Translate’s pricing structure is on a per-character basis, with the first half-million characters falling within the free tier. Google’s pre-trained model is \$20 per million characters, and custom AutoML models are \$80 per million characters.

Google Translate has also developed an active community for the purpose of gathering additional training data with which to refine its models. It uses badges and points to gamify the process, and is integrated into their overall Google contributor framework. Google's contributor framework provides additional incentives and recognition across all of its contributor platforms, such as closer access to Google developers, special events, and early access to prototype versions of new products \cite{ggtrans}.

\subsection{IBM Translation}

IBM Watson Language Translator currently supports 23 languages, although translation between every language pair is not supported. An example would be IBM Watson Language Translators inability to translate from French to Zulu, even though English to Zulu is supported. It supports 21 language-to-English pairs, as well as a select few other language pairs such as French – German, French – Spanish, or Spanish – Catalan. Since the language pairs supported by IBM are much smaller, they are a less viable option for any general language translation framework, unless that framework is specifically looking to go to or from English. One unique feature of IBM’s translation solution is that an instance can be physically placed on premises if necessary, and also all translation data can be withheld from being used in training of IBM’s language pair models. This means that IBM’s approach to translation is much more focused on data security than many of the other solutions, and can thus be utilized in more restrictive environments. 

IBM also supports custom translation models using user-provided word-pair dictionaries or sentence aligned document pairs. Of note for the document pairs, there is only one supported document format, and there are very restrictive criteria such as that there must be at least 5000 pairs and each pair cannot exceed 80 source words. The total corpus size may also not exceed 250 MB.

IBM’s pricing is \$20 per million characters using their standard model, and \$100 per million characters using a custom model. This is one of the more expensive options in the machine translation space \cite{ibmtrans}.

\subsection{Microsoft Translation}

Microsoft's translation currently supports 64 languages, and their service is unique for providing an all-in-one audio-to-translated-text service that can be utilized with a single API call. In order to achieve this, Microsoft developed a tool called Microsoft TrueText \cite{mictt}, which is a pre-processor that normalizes speech-to-text output to prepare it for machine translation. It is unknown if other services such as Google or Amazon perform similar processing internally. Google at least does something similar within their own translate app by removing filler words from speech-to-text (qualitative example: when speaking English into the Google translate app, it doesn’t include any “um” or “ah” filler words. It also predicts between “to”, “too”, and “2”), however it is not clear that there is any sort of processing that would be done in between a Google speech-to-text API call and a Google Translate API call. Microsoft TrueText claims to not only remove filler words and accurately fix close cases based on context, but also to add punctuation and text segmentation as well. One thing to note about TrueText, however, is that since it is used internally, and is not externally usable through an API, it is very hard to test or validate this tool.

It is important to note that any state-of-the-art solution for a speech-to-text to translation pipeline should include some intermediary pre-processing similar to what TrueText is doing in order to improve the accuracy of the final output.

Going back to Microsoft Translate, their API also has the ability to ingest user-provided custom training data to further refine models for specific use cases. This functionality is part of Microsoft's Azure cognitive services ecosystem. In addition to word pair dictionaries, the service is able to ingest entire translated documents as training data, and can even do some internal processing if the two documents are non-sentence aligned. Any user generated models can also be managed and updated via a secure portal or through an API.

Microsoft’s pricing structure is on a per character basis and depends on whether or not a custom model is being utilized. Additionally, their pricing structure has tiers depending on the volume of characters being translated. The free tier includes 2 million characters per month of using any combination of standard and custom models. At the base pay-as-you-go tier, the standard model is priced at \$10 per million characters, and using a custom model is \$40 per million characters. However, tiers for large amounts of characters have further savings on top of these baseline prices making it one of the most affordable options in the translation space. \cite{mictrans}.

\subsection{Vendor Comparison}

\begin{table}[H]
\tabulinesep=1mm
\begin{tabu} to \linewidth{ X[1.2,l,m] X[1,C,m] X[1,C,m] X[1,C,m] X[1,C,m] }
\tabucline[1.5pt]{-}
{} & \textbf{Amazon Translate} & \textbf{Google Cloud Translation} & \textbf{IBM Watson Translator} & \textbf{Microsoft Translation} \\
\hline \hline
\textbf{Base Price} \newline (per million characters) & \$15 & \$20 & \$20 & \$10 \\
\textbf{Free Tier} \newline (characters per month) & 1 million (first 12 months) & 500,000 & 1 million & 2 million \\
\textbf{Custom Training} & \xmark & \cmark & \cmark & \cmark \\
\textbf{Custom Model Price} \newline (per million characters) & N/A & \$80 & \$100 & \$40 \\
\textbf{Languages} & 25 & 104 & 23 & 64 \\
\tabucline[1.5pt]{-}
\tabuphantomline
\end{tabu}
\caption{Translation Vendor Comparison}
\end{table}

\section{Topic Tagging / Multi Label Classification} \label{textsec}
\subsection{BERT}

The field of natural language processing (NLP) has recently been shaken by the emergence of a natural language representation model called Bidirectional Encoder Representations from Transformers, or BERT \cite{bertpaper}. A pre-trained general-purpose BERT model can be fine-tuned to solve many NLP problems with one additional output layer. Use of pre-trained BERT models has led to many substantial improvements in performance across a wide variety of NLP tasks, improving state of the art scores on the GLUE, MultiNLI, SQuAD, SST, and other benchmarks \cite{sota}. Multi label text classification is no exception to this improved performance. Using BERT for this task currently performs the best in comparison to other machine learning models on benchmarking data sets. 
  
The ubiquity of BERT on NLP tasks recently stems from Google's decision to open source BERT. Towards the end of 2018, they released source code developed using TensorFlow, and they built multiple pre-trained representation models. Additionally, BERT has been ported to PyTorch \cite{bertpy}, another deep learning library, and is usable in that ecosystem as well. These developments make it very easy to implement a model on top of BERT without significant overhead (as opposed to training a BERT model from scratch). Because of its availability and performance, any NLP models created going forward should strongly consider leveraging the benefits of BERT \cite{bertgit}.

\subsection{XLNet}

XLNet is an unsupervised natural language representation learning model that is a generalized autoregressive pretraining method for language understanding\cite{xln}. By learning the bi-directional context by using permutations and applying the autoregressive method, XLNet overcomes the pretrain-finetune discrepancy limitation of BERT. The pretrain-finetune discrepancy is a result of the pretraining phase of BERT, the predicted tokens are masked in the input using the symbol [MASK], and [MASK] is not present in real world data during finetuning. As of June 2019, XLNet outperforms BERT on 20 NLP tasks and achieves state of the art results on 18 of the tasks, such as sentiment analysis, multilabel text classification, question answering, document ranking, and natural language inference\cite{xldown}. XLNet is performing the best in comparison to BERT and other machine learning models on benchmarking data sets in tasks.

Example enrichment from the example for each task is available below:
\begin{itemize}
   \item Topics and Tags: \textbf{\textit{Economy, Trade}}
   \item Named Entities: \textbf{\textit{Mr. K, Mrs. K, Dataland, ACME, Umbrella Co.}}
   \item Sentiment Analysis: \textbf{Positive}
   \item Question Answering: Who is Umbrella Co. working with? \textbf{ACME}
   \item Summarization: \textit{"ACME and Umbrella Co. have agreed on a deal to reduce shipping costs."}
\end{itemize}

\section{Object Recognition}

Object Recognition is the task of utilizing machine learning to find objects within an image or video. This can be used to identify faces, animals, scenes or environments, text, or nearly any other object. This process is enhanced by the feature detection capabilities of machine learning and deep learning, which doesn't require any pre-processing (such as image kerneling). This process is completed by most software in the following three step process: discovery, correct identification, and labeling. In their current state, most out-of-the-box Object Recognition services work with real time video, provide explicit content detection, celebrity recognition, face detection, include Optical Character Recognition (OCR), and are being continually improved. 

Example Information that could be detected (Fig. \ref{news_or}):
\begin{figure}[H]
  \centering
  \includegraphics[width=13cm]{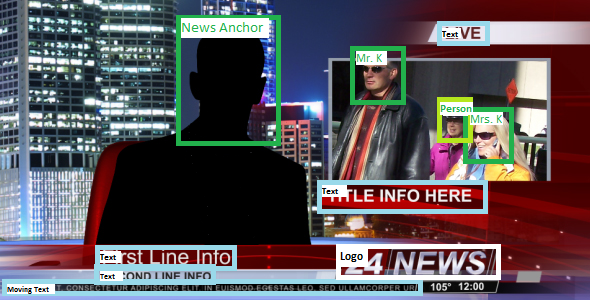}
  \caption{Sample Bounding Boxes}
  \label{news_or} 
\end{figure}

\subsection{Amazon Rekognition}

Amazon Rekognition provides a wide variety of services using object recognition technology developed by their vision scientists. A key feature of Rekognition is not only face detection, but face analysis, which includes recognizing a wide range of emotions and moods. In our example, this could detect that Mrs. K is happy, thus supporting the sentiment analysis that the segment on the deal was positive. Pathing for objects is another useful feature included with Amazon Rekognition. This means the software can track the location of an object over a period of time. Model training, as with most Amazon products, is limited. It has the capability to accept custom labels for object detection, and characters for text in image, but extensive training of provided models is limited. Amazon Rekognition supports JPEG, PNG, H.264, MPEG-4, and MOV formats; with SDKs in JavaScript, IOS, Android, Java, .NET, Node.js, PHP, Ruby, and Python. Their service works with a batch size up to 8GB or 2 hours of video, and meta data is included in the output. However, Amazon Rekognition does require use of their S3 network to process videos.  

Rekognition pricing begins at \$0.10 per minute, with reductions for large amounts of video and a free trial available.  But these small figures can quickly add up \cite{amzor}.

\subsection{Google Video AI}

Google offers two Video AI products for different use cases. AutoML Video Intelligence is a Video analysis tool that makes it easy to train custom models and includes an easy to use graphical interface. Google's Video Intelligence API has a collection of pre-trained models that are optimized for common use cases. The Intelligence API supports MOV, MPEG4, MP4, and AVI formats; both versions use REST and RPC APIs allowing for easy implementation. Video processing features include shot detection, extensive transcription capabilities, keyword detection, OCR features, and valuable meta data.    

AutoML video has a free version and sports pricing per node, while Intelligence API also has a free version, and offers prices for individual features, dependent on stored and streamed video.  

Google Video Intelligence's pricing scheme varies based on what information is extracted from the video, and can be inexpensive for simple tasks, but quickly add up for a full feature set \cite{ggor}.

\subsection{Microsoft Video Indexer}

Microsoft Video indexer is one of the more versatile Object Recognition services on the market. Sold as an all in one package, it sports an impressive set of features, including thumbnail extraction for faces, shot detection, keyframe and black frame detection, credit detection, audio transcription, OCR services, two channel processing, noise reduction, speaker enumeration, audio effects and emotion detection translation. It also includes useful meta data like speaker statistics, automatic language detection, and account-based face identification. One of the more rare and applicable traits of Microsoft Video Indexer is its scene identification qualities. This allows a user to have automate the process of finding parts of a video specific to their needs. Google also employs a form of segmentation in their software, comparable to Microsoft’s. Most video types are accepted and JSON is the standard return format.
   
Free services are available, with individual prices varying by video quality \cite{micor}.

\subsection{Vendor Comparison}

\begin{table}[H]
\tabulinesep=1mm
\begin{tabu} to \linewidth{ X[1.2,l,m] X[1,C,m] X[1,C,m] X[1,C,m] }
\tabucline[1.5pt]{-}
{} & \textbf{Amazon Rekognition} & \textbf{Google Video AI} & \textbf{Microsoft Video Indexer} \\
\hline \hline
\textbf{Base Price} (per minute) & \$0.10 & \$0.10 - \$0.50 & \$0.12 \\
\textbf{Free Tier} (minutes) & 1000 (first 12 months) & 1000 & 600 (basic), 2400 (API) \\
\textbf{Custom Models} & \cmark & \cmark & \cmark \\
\textbf{Batch Size} (GB) & 8 & 10 & 10 \\
\textbf{Accepted Formats} & JPEG, PNG, H.264, MPEG-4 & H.264, MPEG-4 & Not specified \\
\tabucline[1.5pt]{-}
\tabuphantomline
\end{tabu}
\caption{Object Recogniton Vendor Comparison}
\end{table}

\subsection{Supplementary Options}
\subsubsection{Yolo (You Only Look Once)}

YOLO is an open source object recognition product that uses a Fully Convolutional Neural Network (FCNN) with 24 convolutional layers followed by 2 fully connected layers. Running at 45 frames per second on their base version, with faster more compact free versions available, YOLO is one the most accurate real time Object recognition products on the market, processing video in real time with less than 25 milliseconds of latency. Pre-trained models are provided on their website and through other sources such as GitHub, although their home-grown version is not currently marketed for business use.  YOLO does tend to struggle when small objects in groups (such as a flock of birds) and its main source of error is incorrect localization. YOLO also requires extensive coding to train and work with \cite{yolo}.

\subsubsection{Mask R-CNN}

Mask R-CNN is an object detection algorithm that simultaneously detects objects in an image or video frames with bounding box recognition and generates a mask over the object area.\cite{maskcn} Mask R-CNN is an extension of Faster R-CNN, where for each object output there is a class label and bounding box, with an addition of the object mask. If using a single model approach over an ensemble model approach, Mask R-CNN outperforms object detection algorithms in single model use. Mask R-CNN has pretrained models implemented on PyTorch, Keras, and Tensorflow. Mask R-CNN can serve as a solid baseline to many object detection tasks. For details on Mask R-CNN results and research, visit Facebook AI Research’s objection detection github page at https://github.com/facebookresearch/Detectron. 

\section{Conclusion}

Video intelligence has recently become a critical field for anyone seeking to uncover data in the vast digital world. The AI-based analytics and automation tools presented in this paper are the best in class at extracting and structuring video based data. The latest and greatest in machine learning technology, machine transcription, machine translation, topic tagging, and object recognition tasks are improving constantly, breaking performance records in speed and accuracy as new applications evolve. Therefore, more than ever, it is necessary to stay up data on recent advances in video intelligence. 

Machine transcription plays on of the strongest roles in video analytics, since the majority of video data originates as speech. All the services reviewed provide the needed features to properly transcript audio data, leaving the deciding factors to be cost and additional features. Based on price alone, Google Cloud Speech-to-Text wins out at only \$0.016 per minute for their base model. Although, they do sell additional features separately, with more advanced models priced around \$0.024 per minute. This opens the door for supplementary open source software such as Kaldi or Mozilla \& DeepSpeech. As with any open source software, being free comes at a cost, which in this case means having to train a model from the ground up. As for ease of use and out of the box efficiency, Amazon Transcribe performs best. With their all in one price of \$0.024 everything comes included, from simple yet effective implementation features to high levels of accuracy across multiple video types. For anything language intensive, Google Cloud Speech-to-Text pulls from Google's extensive library of translated datasets, allowing for over 120 transcriptable languages. IBM Watson Speech-to-Text provides the most customizable, pre-trained transcription engine, but keep in mind that anything open source will also be highly customizable. Finally, for more personal use transcription software, Nuance's Dragon Speech Recognition Software excels. Their model must be trained specifically to the users voice and tone, with a minimum sampling rate of 16 kHz.

The variety of countries with a strong virtual footprint has increased dramatically in the past decade. And recent advancements in machine learning technology have allowed for efficient translation provided by implementing post processing layers. In regards to price, Microsoft Translation pushes ahead. Priced at \$10 per million characters, they also provide an extensive free version that allows for 2 million free characters every month. Both Amazon and Google perform exceptionally out of the box, supporting high levels of functionality and accuracy. Unsurprisingly, Google has the most languages available, sporting 104 fully functional languages. IBM and Microsoft Translation win out for customization features, even allowing you to feed entire documents to their system as training data. The reason we did not include open source services, is due to the difficulty of training a translation model, with all its required languages, from the ground up.

For Text Classification, we recommend fine-tuning a BERT model for NLP tasks as it will allow for high-accuracy and processing of as much data as your machine would handle. In regards to buying or building a machine, it may be expensive upfront, as it is best to train BERT with GPU's such as NVIDIA Turing Series GPU's that are optimized for machine learning, but the accuracy and long-term costs led us to determine it as the best option. One caveat would be the release and success of a more recent text Classification model known as XLNet. Outperforming BERT in more than 20 NLP tasks, XLNet has proven itself to be a formidable opponent in the highly competitive NLP market. 

Object Recognition has proven itself to be a more difficult category to survey due to the task overlap seen in many services. An example of this would be the transcription capabilities included in Google Video AI. Their video recognition software can automatically transcribe audio, but only in English, even though their stand alone transcription service works with over 120 languages. The most cost effective service is Amazon Rekognition. Formatted to contain a variety of high quality recognition features it clocks in at \$0.10 per minute, with a free tier included. Google Video AI and Microsoft Video Indexer both implement transcription features into their recognition software, and Microsoft Video Indexer even allows for translation. All three services allow for a range of customization, although YOLO, the supplementary open source service, grants ground up training. These services also all incorporate Optical Character Recognition (OCR) to enhance video analytics. 

At the current stage, there isn't a clear overall recommendation for a single solution for video analytics, but rather it varies by use-case, scope and available resources. For our example of enriching worldwide news sources, we expect significant investment and computational resources will be required. International news collection will have the biggest challenges in the realm of transcription and translation, as accuracy isn't consistent for all languages, nor do all providers have the same pool of languages. For this scenario we'd recommend Google's Cloud services for the first two stages, using their pre-trained models and occasionally adding custom dictionaries. This decision is based off their prices for transcription and translation, which is among the lowest, and the vast support for different languages. One feature that is particularly useful for news is facial recognition, which is why we lean towards a vendor like Amazon, rather than open source. In the end both Amazon and Google have similar features and pricing, so the choice between the two would depend on the environment of a business and what service API's are most familiar.

Alternatively, for smaller scale applications - such as in-house video processing of corporate internal seminars - open source solutions are likely capable of handling all the data with cost being affordable. Most costs will be upfront in the purchasing, setup, and maintenance of HPC's, as well as time needed to train and produce the data for training the models. Transcription through Kaldi or DeepSpeech frameworks could be trained in a way that internal jargon is transcribed efficiently, and errors are lower. The translation step would either be minimal or non-existent, and if needed Microsoft's prices are worth considering despite a smaller pool of languages than Google. Additionally, Microsoft Video Indexer combines transcription, translation, OCR, and object recognition into one package, specialized for easy deployment through Azure; which could be a viable choice for a smaller scale application. Finally BERT for text analysis and YOLO for video analysis would be easy to train and get accurate results, with low long-term costs for the company.

After evaluating the state of the art in video enrichment and extraction services, we see a need for a methodology to aggregate these disparate services into a streamlined process to develop more comprehensive video intelligence (VIDINT). In order to fully exploit advances in each sub-task a VIDINT system must include four key machine learning-based functions: 1) aggregation, 2) video selection, 3) task-based service optimization and 4) summarization. The recent release of Youtube's 8M dataset\cite{8m} is perfect for training such a system, since the 350,000 hours of data provided by it comes pre-processed and segmented. Development of an integrated machine learning system leveraging the tools from this survey is the subject of our current work. 

\printbibliography
\end{document}